\documentclass[aps,prl,epsfig,showpacs,floats,twocolumn,superscriptaddress,amssymb,amsmath,floatfix]{revtex4}
\usepackage{graphicx}
\begin{document}

%
\def\f#1#2{\frac{#1}{#2}}
\def\der#1#2{\f{\D #1}{\D #2}}
\def\fder#1#2{\f{\delta #1}{\delta #2}}
\def\D{\partial}
\def\grad{\nabla}
\def\div{\nabla \cdot}
\def\rot{\nabla \times}
\def\lap{\nabla^2}
\def\inv{^{-1}}
\def\Tr{\mbox{\rm Tr}\;}
\def\const{\mbox{const}}
\def\etal{{\it et al.}}
\def\Av#1{\overline{#1}}
\def\Ref#1{(\ref{#1})}
\def\Eq#1{Eq.(\ref{#1})}
\def\Eqs#1{Eqs.(\ref{#1})}
\def\Ref#1{(\ref{#1})}
\def\Fig#1{Fig.\ref{#1}}
\def\dis{\displaystyle}
\def\eq{\begin{eqnarray}}
\def\qe{\end{eqnarray}}
\def\eqnn{\begin{eqnarray*}}
\def\qenn{\end{eqnarray*}}
\def\nn{\nonumber}
\def\bA{\bm{A}}
\def\bB{\bm{B}}
\def\bC{\bm{C}}
\def\bD{\bm{D}}
\def\bE{\bm{E}}
\def\bF{\bm{F}}
\def\bG{\bm{G}}
\def\bH{\bm{H}}
\def\bI{\bm{I}}
\def\bJ{\bm{J}}
\def\bK{\bm{K}}
\def\bL{\bm{L}}
\def\bM{\bm{M}}
\def\bN{\bm{N}}
\def\bO{\bm{O}}
\def\bP{\bm{P}}
\def\bQ{\bm{Q}}
\def\bR{\bm{R}}
\def\bS{\bm{S}}
\def\bT{\bm{T}}
\def\bU{\bm{U}}
\def\bV{\bm{V}}
\def\bW{\bm{W}}
\def\bX{\bm{X}}
\def\bY{\bm{Y}}
\def\bZ{\bm{Z}}
\def\ba{\bm{a}}
\def\bb{\bm{b}}
\def\bc{\bm{c}}
\def\bd{\bm{d}}
\def\be{\bm{e}}
\def\beff{\bm{f}}
\def\bg{\bm{g}}
\def\bh{\bm{h}}
\def\bi{\bm{i}}
\def\bj{\bm{j}}
\def\bk{\bm{k}}
\def\bl{\bm{l}}
\def\bem{\bm{m}}
\def\bn{\bm{n}}
\def\bo{\bm{o}}
\def\bp{\bm{p}}
\def\bq{\bm{q}}
\def\br{\bm{r}}
\def\bs{\bm{s}}
\def\bt{\bm{t}}
\def\bu{\bm{u}}
\def\bv{\bm{v}}
\def\bx{\bm{x}}
\def\by{\bm{y}}
\def\bz{\bm{z}}
\def\bA{\bm{A}}
\def\bB{\bm{B}}
\def\bC{\bm{C}}
\def\bD{\bm{D}}
\def\bE{\bm{E}}
\def\bF{\bm{F}}
\def\bH{\bm{H}}
\def\bI{\bm{I}}
\def\bJ{\bm{J}}
\def\bK{\bm{K}}
\def\bL{\bm{L}}
\def\bM{\bm{M}}
\def\bN{\bm{N}}
\def\bO{\bm{O}}
\def\bP{\bm{P}}
\def\bQ{\bm{Q}}
\def\bR{\bm{R}}
\def\bS{\bm{S}}
\def\bT{\bm{T}}
\def\bU{\bm{U}}
\def\bV{\bm{V}}
\def\bW{\bm{W}}
\def\bX{\bm{X}}
\def\bY{\bm{Y}}
\def\bZ{\bm{Z}}
\def\ten{$\bullet$ \,}
\def\nten{\noindent $\bullet$ \,}
\def\vsp#1{\vspace{#1 mm}}
\def\hsp#1{\hspace{#1 mm}}
\def\simge{\;\lower3pt\hbox{$\stackrel{\textstyle >}{\sim}$}\;}
\def\simle{\;\lower3pt\hbox{$\stackrel{\textstyle <}{\sim}$}\;}
\def\bm#1{\mbox{\boldmath $#1$}}
\def\tensor#1{{\sf #1}}
\def\lrL#1{\left[#1\right]}
\def\lrM#1{\left\{#1\right\}}
\def\lrS#1{\left(#1\right)}
\def\lrF#1{\left|#1\right|}
\def\lrA#1{\left\langle #1 \right\rangle}
\def\lrAA#1{\left\langle\!\left\langle #1 \right\rangle\!\right\rangle}
\def\biglrL#1{\bigl[ #1 \bigr]}
\def\biglrM#1{\bigl\{#1 \bigr\}}
\def\biglrS#1{\bigl( #1 \bigr)}
\def\biglrF#1{\bigl|#1\bigr|}
\def\biglrA#1{\bigl\langle #1 \bigr\rangle}
\def\bigglrL#1{\biggl[ #1 \biggr]}
\def\bigglrM#1{\biggl\{#1 \biggr\}}
\def\bigglrS#1{\biggl( #1 \biggr)}
\def\bigglrF#1{\biggl|#1\biggr|}
\def\bigglrA#1{\biggl\langle #1 \biggr\rangle}
\def\BigglrL#1{\Biggl[ #1 \Biggr]}
\def\BigglrM#1{\Biggl\{#1 \Biggr\}}
\def\BigglrS#1{\Biggl( #1 \Biggr)}
\def\BigglrF#1{\Biggl|#1\Biggr|}
\def\BigglrA#1{\Biggl\langle #1 \Biggr\rangle}
\def\commentoff#1{}
\def\commenton#1{{\sf #1}}
%
\renewcommand{\br}{{\bf r}}
\renewcommand{\bR}{{\bf R}}
\renewcommand{\bx}{{\bf x}}
\renewcommand{\bv}{{\bf v}}
\renewcommand{\bG}{{\bf G}}
\renewcommand{\bH}{{\bf H}}

\newcommand{\tomega}{\Omega}


\title{Synchronization in A Carpet of Hydrodynamically Coupled Rotors
with Random Intrinsic Frequency}

\author{Nariya Uchida}
\email{uchida@cmpt.phys.tohoku.ac.jp}
\affiliation{Department of Physics, Tohoku University, Sendai, 980-8578, Japan}

\author{Ramin Golestanian}
\email{r.golestanian@sheffield.ac.uk}
\affiliation{Department of
Physics and Astronomy, University of Sheffield, Sheffield S3 7RH, UK}


\date{\today}

\begin{abstract}
We investigate
synchronization caused by long-range hydrodynamic interaction
in a two-dimensional, substrated array of rotors
with random intrinsic frequencies.
The rotor mimics a flagellated bacterium that is
attached to the substrate (``bacterial carpet'')
and exerts an active force on the fluid.
Transition from coherent to incoherent regimes is
studied numerically, and the results are compared to
a mean-field theory.
We show that quite a narrow distribution of the intrinsic
frequency is required to achieve collective motion
in realistic cases.
The transition is gradual, and the critical behavior 
is qualitatively different from that of the conventional 
globally coupled oscillators.
The model not only serves as a novel example of non-locally coupled
oscillators, but also provides insights into the role of
intrinsic heterogeneities in living and artificial
microfluidic actuators.
\end{abstract}

\maketitle

\paragraph{Introduction. -}

Collective oscillations of active elements
are observed
in a variety of
physical, chemical, and biological systems far from equilibrium.
Numerous studies have been devoted to the mutual entrainment of
oscillators that have different intrinsic frequencies~\cite{Kuramoto,Acebron}.
A class of phase oscillators with global (or mean-field)
coupling have enjoyed deep theoretical understanding~\cite{Kuramoto,Acebron,Crawford},
while a myriad of unresolved problems still remain on
the behaviors of locally~\cite{Sakaguchi} and non-locally~\cite{Kuramoto95,Strogatz}
coupled oscillators.
In particular, knowledge about synchronization caused by
long-range interactions is quite limited~\cite{Rogers,Marodi,Zaslavsky},
although they are ubiquitous in Nature in the form of,
e.g., gravitational, electromagnetic, elastic, and hydrodynamic forces.

Biologically important examples of
long-ranged synchronization are provided by
swimming microorganisms that are interacting
hydrodynamically, such as sperm flagella beating in
harmony~\cite{Taylor,Kruse,holger,GompperPRE,RaySync}, and 
metachronal waves in cilia~\cite{metachron,lagomarsino,Lenz,vilfan,Joanny}.
Both of these oscillatory elements, flagella and cilia, are
driven by molecular motors embedded in the cell surface,
and interact through the viscous environment (water).
In order to describe their synchronized motion,
several theoretical models have been proposed~\cite{Taylor,GompperPRE,metachron,lagomarsino,holger,Lenz,vilfan,Joanny}.
While all of these initial studies are focused on a homogeneous set of identical systems, it will be important to consider the role of disorder, as real biological motors possess
intrinsic heterogeneities that could affect the collective dynamics.

An example of collective yet heterogeneous dynamics
is found in a bacterial carpet, which is recently
introduced as a new type of microfluidic device~\cite{Berg}.
The assembly is composed of a dense monolayer of bacteria that are
attached to a solid substrate by their bodies (heads).
Their flagella (tails), on the other hand, can freely rotate in the fluid
and are orientationally ordered by hydrodynamic interaction,
to generate coordinated fluid motion.
Evolution of correlated regions is observed, but
the ordering remains partial.
Observations of irregular and slowly varying
flow structures ('whirlpools' and 'rivers')
suggest the presence of heterogeneity in the configurations
of the rotors~\cite{KimBreuer} . Fabrication of more efficient
microfluidic pumps could be achieved through understanding
and controlling the heterogeneity.

Recently, we have proposed a generic model of
hydrodynamically coupled rotors arrayed on a substrate,
and studied the collective dynamics of uniform elements~\cite{hsync-paper1}.
In this letter, using a variant of the model,
we address the effect of random intrinsic frequencies
on synchronization.
To be concrete, we consider a simple and idealized model
of bacterial carpets, where a flagellated bacterium is mounted
to each rotor nearly (but not completely) radially.
Depending on the mounting angle of the flagellum,
it generates a torque that drives the rotor in either
clockwise or counterclockwise direction.
We assume that the mounting angle
and hence the intrinsic frequency of the rotor
are randomly distributed.
By varying the degree of randomness, we study the transition
from coherent to incoherent regimes numerically.
Our results suggest that synchronization of the rotors,
and hence collective pumping of the fluid, requires
a quite narrow distribution of the mounting angle
in realistic cases.
In order to understand the transition behavior,
we apply the mean-field theory, which is originally developed
for global coupling, to our long-ranged system.
While we obtain a fair agreement between theory and simulation
for the synchronization threshold, the transition is shown
to be more gradual than in globally coupled oscillators.

\paragraph{Model. -}

\begin{figure}
\includegraphics[width=0.6\columnwidth]{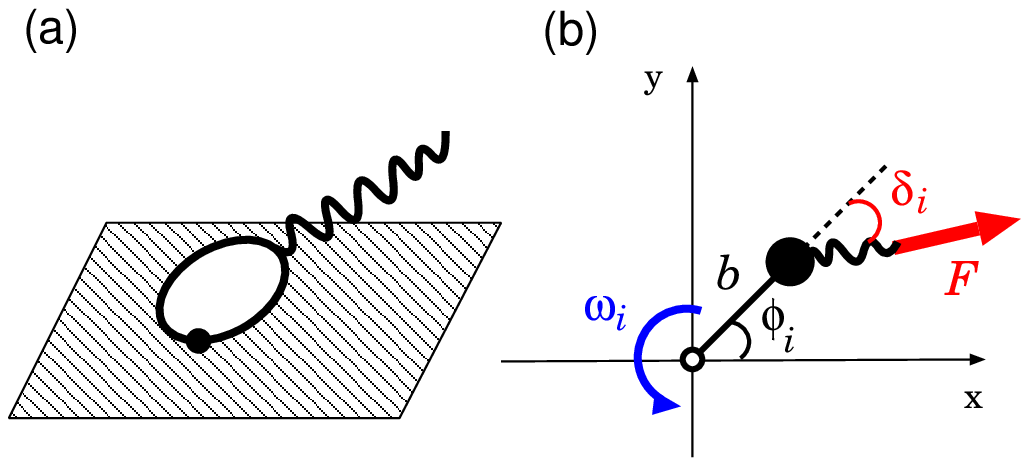}
\caption{Schematic picture of the rotor (top view).
The active force ${\bF_i}$ exerted by the $i$-th rotor
on the fluid is deviated by a fixed angle $\delta_i$
from the radial direction. The reaction force drives the rotor
at the intrinsic frequency $\omega_i =F\sin \delta_i/\zeta b$,
where $b$ is the radius of rotation and $\zeta$ the viscous drag coefficient
of the bead.
}
\label{schematic}
\end{figure}

We consider an array of rotors positioned on a square lattice
of grid size $d$.
Each rotor has a thin, freely-rotatable arm on the tip
of which a flagellated bacterium is mounted.
The bacterium consists of a spherical bead of radius $a$ (body)
and a thin tail that lies horizontally (flagellum).
Motion of the bead is constrained on a circular orbit of radius $b$
located at height $h$ from the substrate, which we take to be
the $xy$-plane.
The position of the $i$-th bead is thus given by
$\br_i = \br_{0i} + h \be_z + b \bn_i$ where
$\br_{0i}$ is its base position on the square lattice
and
$\bn_i= (\cos\phi_i, \sin \phi_i, 0)$ is the unit vector
that gives the orientation of the arm
via its phase $\phi_i = \phi_i(t)$.
The velocity of the bead reads $\bv_i = \dot{\phi}_i \bt_i$
where $\bt_i = (-\sin\phi_i, \cos\phi_i, 0)$ is
the unit vector tangential to the trajectory.
We assume that the active force $\bF_i$ exerted by
the rotor on the fluid has a constant magnitude $F$,
and makes a fixed angle $\delta_i$ (measured clockwise)
from the radial direction;
$\bF_i = F (\cos \delta_i \bn_i - \sin \delta_i \bt_i)$.
See Fig. \ref{schematic} for the configuration.
The reaction force $-\bF_i$ on the rotor arm gives
the driving torque $T_i= F b \sin \delta_i$ and the
intrinsic frequency $\omega_i = F \sin \delta_i/\zeta b$,
where $\zeta = 6\pi \eta a$ is the viscous drag coefficient.
The mounting angles $\delta_{i}$'s are
assumed to have the Gaussian distribution
\eq
P(\delta) = \f{1}{\sqrt{2\pi} \delta_0}
\exp\left(-\f{\delta^2}{2\delta_0^2}\right)
\qe
with the standard deviation $\delta_0$.

We assume that the rotors are widely spaced so that $a, b, h \ll d$.
Then the velocity field of the fluid created by the active forces
is given by $\bv(\br) = \sum_i {\bf G}(\br - \br_i) \cdot \bF_i$,
where
${\bf G}(\br) = (3h^2/2\pi \eta) \cdot \br_\perp \br_\perp/|\br|^5$,
$\br_\perp = (x,y,0)$ is the asymptotic expression of the
Oseen-Blake tensor~\cite{Blake} for $h/d \ll 1$.
The rotor's angular velocity is given by
$\omega_i + \bv(\br_i) \cdot \bt_i /b$,
or, more explicitly,
\begin{eqnarray}
\frac{d \phi_i}{d t} &=& \omega_0 \sin \delta_i -
\frac{3 \gamma \omega_0 d^3}{4\pi}
\sum_{j\neq i}
\frac{1}{|{\bf r}_{ij}|^3}
\Bigg[
\sin(\phi_i - \phi_j + \delta_j)
\nonumber\\
& & + \cos(\phi_i + \phi_j - \delta_j - 2\theta_{ij})
\Bigg].
\label{dotphi1}
\end{eqnarray}
Here,
$\omega_0 = F/\zeta b$, 
${\bf r}_{ij} = {\bf r}_i - {\bf r}_j
= |\br_{ij}|(\cos \theta_{ij}, \sin \theta_{ij})$,
and
$\gamma= \zeta h^2/\eta d^3 = 6\pi a h^2/d^3$
is the dimensionless coupling constant.
For a real bacterial carpet, $a\sim h \sim 1 \mu$m and
$d \sim 10\mu$m give the rough estimate $\gamma \sim 10^{-2}$.
\par

\paragraph{Numerical Simulation. -}

We implemented the model on a $L \times L$ square lattice
and numerically integrated eq. (\ref{dotphi1}) by the Euler method.
We assumed the periodic boundary condition and computed
the velocity field every time step in the Fourier space.
We set $\gamma=0.1$ and varied the angle deviation $\delta_0$
as the control parameter.
The system size used was $L=128$ for most of the results shown below,
while $L=32$, $64$ and $256$ are also used to check finite-size effect.
Starting from random initial configurations of $\phi_i(t=0)$,
the system reached a dynamical steady state by the time
$t=1 \times 10^4/\omega_0$.
The statistical data shown below are taken from 
the time window $1 \times 10^4 < \omega_0 t < 2.5 \times 10^5$.

\begin{figure}
\includegraphics[width=0.9\columnwidth]{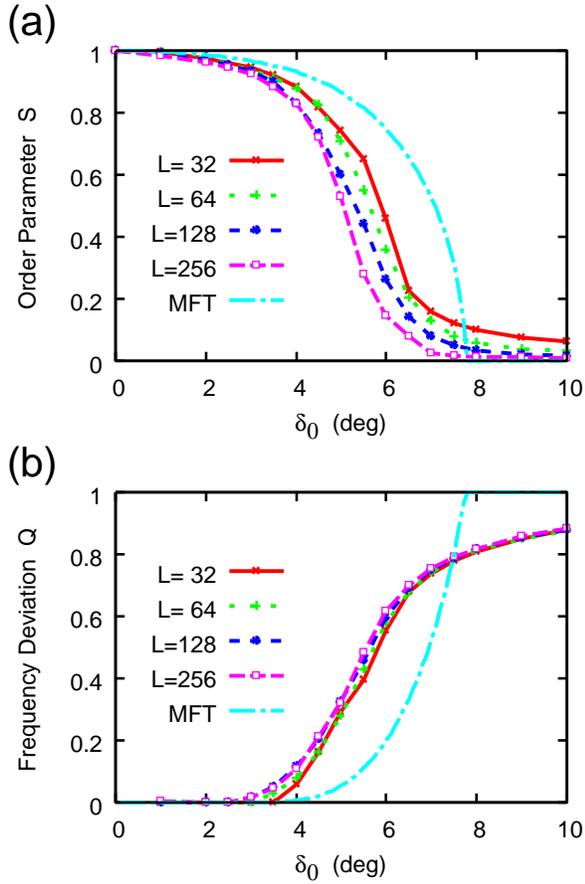}
\caption{
Transition behavior for system size $L= 32, 64, 128$, and 256
with comparison to the mean-field theory (MFT).
(a)
Orientational order parameter
$S=|\langle \bn \rangle| = |\langle e^{i\phi} \rangle|$
versus $\delta_0$.
(b) Normalized frequency deviation
$Q = \sqrt{\langle \tomega^2 \rangle/
\langle {\omega}^2 \rangle}$ versus $\delta_0$
versus $\delta_0$.
}
\label{orderparam}
\end{figure}

We plot the orientational order parameter
$S=|\langle \bn \rangle| = |\langle e^{i\phi} \rangle|$
as a function of $\delta_0$ in Fig. \ref{orderparam}.
Also shown is
the standard deviation (STD) of the actual frequency
$\tomega_i = \langle\dot{\phi_i}\rangle$
normalized by the STD of the intrinsic frequency $\omega_i$,
$Q = \sqrt{\langle \tomega^2 \rangle/\langle {\omega}^2 \rangle}$.
Note that $S=1$ and $Q=0$ in the fully synchronized state
and $S=0$ and $Q=1$ in the desynchronized limit.
As we increase $\delta_0$, $S$ and $Q$ slowly converge to
the desynchronized limit.
While the change in the orientational order parameter is
sharper for a larger system size, the frequency deviation
has little $L$-dependence.
For comparison, we also show the results of the mean-field theory,
which will be explained in the next section.

\begin{figure}
\includegraphics[width=0.9\columnwidth]{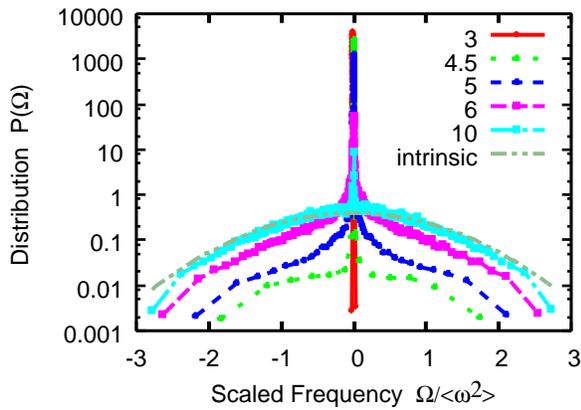} 
\caption{
Distribution of the normalized actual frequency
$\tomega/\sqrt{\langle {\omega}^2 \rangle}$
as functions of $\delta_0$.
For comparison, the distribution of the intrinsic frequency
$\omega_i$ is also shown.
}
\label{frequency}
\end{figure}

In Fig. \ref{frequency}, we plot the distribution function
of the actual frequency $\tomega$
normalized by the STD of intrinsic frequency,
for different values of $\delta_0$.
The distribution consists of a sharp delta-function like peak
at $\tomega=0$ and broad symmetric tails for
$\tomega>0$ and $\tomega<0$.
For $\delta_0 \le 3^\circ$,
most of the rotors are coherent and contribute to the center-peak.
For $\delta_0=10^\circ$, the distribution is close to that of
the intrinsic frequency, while the center peak still remains.
the above data suggest that the synchronization transition in this system is
more gradual than that found in a globally coupled system,
and it is difficult to locate the transition point exactly.

\begin{figure}
\includegraphics[width=0.9\columnwidth]{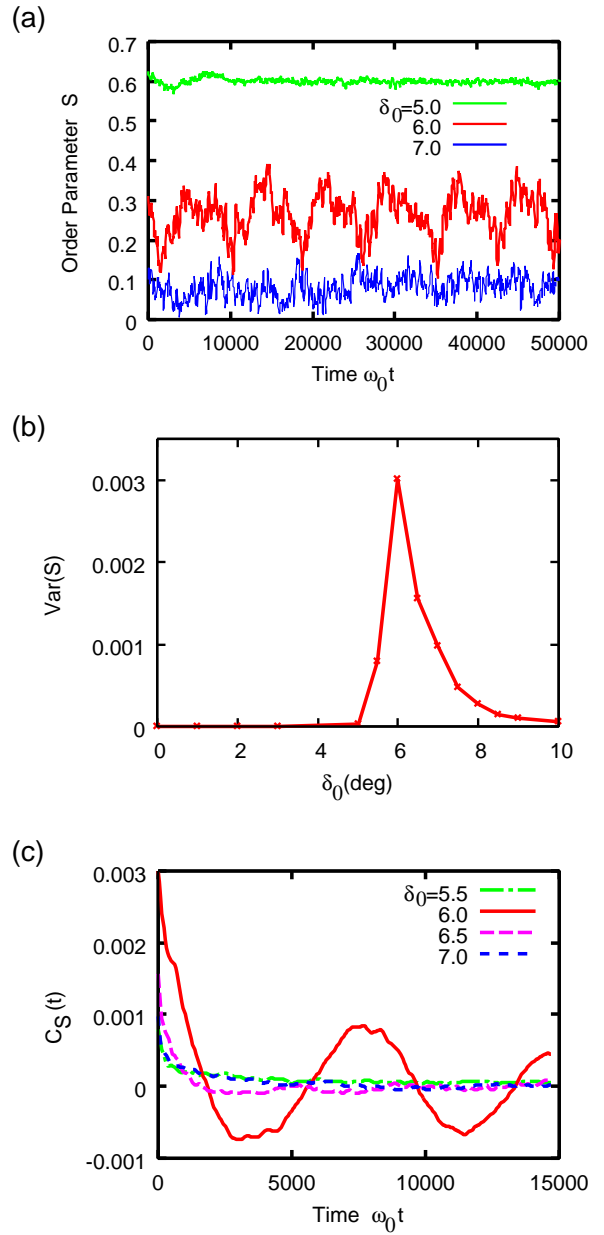}
\caption{
(a)
Variance of the order parameter Var$(S)$ as a function of $\delta_0$.
Fluctuation is most enhanced at $\delta_0 = 6^\circ$.
(b)
Auto-correlation function of the order parameter $C_S(t)$.
An oscillatory behavior is prominent at $\delta_0 = 6^\circ$.
(c)
Time series $S(t)$.
}
\label{variance}
\end{figure}

On the other hand, the variance of the order parameter
Var$(S) = \langle S(t)^2 \rangle - \langle S(t) \rangle^2$
as a function of $\delta_0$ (Fig. \ref{variance}(a))
has a peak near $\delta_0= 6^\circ$,
suggesting that there is a subtle balance between
synchronization and desynchronization.
We will call this the threshold angle
and denote by $\delta_{th}$.

In Fig.\ref{variance}(b), we plot the temporal
correlation function of the order parameter
$C_S(t)= \langle S(t+t')S(t')\rangle - \langle S \rangle^2$.
We find an oscillatory behavior with long correlation time
at $\delta_0 = 6^\circ$.
The presence of a characteristic period is also directly
observed in the plot of $S(t)$ in Fig. \ref{variance}(c).
Although the origin of the oscillation is beyond the scope of
the present paper, a preliminary study shows that the oscillatory
behavior at the threshold angle is a unique feature resulting
from the long-ranged nature of the interaction.

\begin{figure}
\includegraphics[width=0.9\columnwidth]{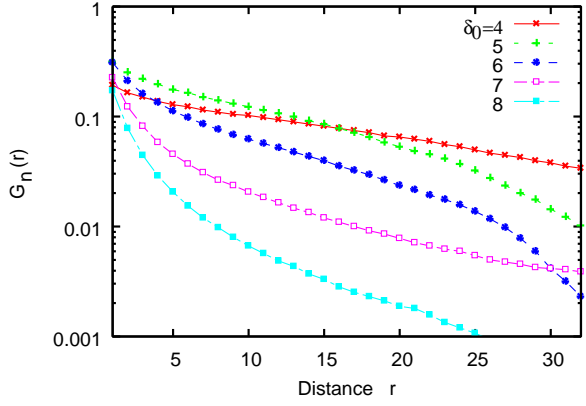}
\caption{
Orientational correlation function $G_n(r)$.
Nearly exponential decay of correlation is observed
for $\delta_0 \le 6^\circ$.
}
\label{correlation}
\end{figure}

We also plot the orientational correlation function
\begin{equation}
G_n(|\br|) =
\langle
[\bn(\br+\br') - \overline{\bn}] \cdot
[\bn(    \br') - \overline{\bn}]
\rangle
\end{equation}
in Fig.\ref{correlation}.
Here the angular brackets mean taking
average over $\br'$ as well as the azimuthal angle of $\br$.
For $4 \le \delta_0 \le 6^\circ$,
we observe an exponential decay of the correlation
over a wide distance.
For $\delta_0 > 6^\circ$, on the other hand, the correlation
is short-ranged and decays more slowly than exponential.
The qualitative change in the correlation function
gives another support of the above estimate of
the threshold angle, $\delta_{th}=6^\circ$.

\paragraph{Mean-field theory. -}
Now we apply a mean-field approximation to our model.
When the coupling is weak ($\gamma \ll 1$)
and the force angle is small ($\delta_0\ll 1$),
we can  neglect the $\delta_j$'s in the RHS of eq. (\ref{dotphi1})
because they give $O(\tilde{\gamma} \delta_0)$ contributions.
We also replace the interaction kernel $\propto \br\br/r^5$
by its angular average,  as a result of which
the cosine term in the RHS of eq. (\ref{dotphi1}) is dropped.
This gives the phase equation in the familiar form,
\begin{equation}
\frac{d \phi_i}{d t}= \omega_i -
\sum_{j\neq i} G({\bf r}_i - {\bf r}_j) \sin(\phi_i - \phi_j)
\label{dotphi2}
\end{equation}
with $\omega_i = \omega_0 \sin \delta_i$ and
$G({\bf r}) = 3 \gamma \omega_0 d^3/4\pi r^3$.
The distribution of the intrinsic frequency is
given by $P_\omega(\omega) = |d \delta/d\omega| P(\delta)
= P(\sin^{-1}(\omega/\omega_0))/\sqrt{\omega_0^2 - \omega^2}$.
Now we apply the mean-field ansatz which was originally 
proposed by Kuramoto~\cite{Kuramoto} for global coupling: 
\eq
R e^{i\theta}= \sum_j G(\br_i-\br_j) e^{i\phi_j}.
\label{ansatz}
\qe
Here, the amplitude $R$ and the phase $\theta$
of the order parameter are assumed to be constant
in space and time, which would be justified if
the interaction is sufficiently long-ranged.
The isotropy of $G(\br)$ allows us
to assume $\theta=0$ without loss of generality.
Using this we can rewrite eq. (1) as
\eq
\f{d\phi_i}{dt} &=& \omega_i - R \sin \phi_i.
\qe

Equation (3) allows a stationary ($\dot{\phi}=0$) solution
if and only if $|\omega_i| < R$.
The rotors satisfying this condition
have the actual frequency $\tomega=0$
and are called the coherent group.
The phase of a rotor belonging to this group is given by
\eq
\phi_i = \sin^{-1} \left( \f{\omega_i}{R} \right).
\label{phi_coh}
\qe
where the principal value of the inverse-sine function
should be chosen so that $|\phi| < \pi/2$.
The phase distribution $n(\phi)$ of the coherent group reads
\eq
n(\phi)
= P_\omega(\omega) \cdot \left| \f{d\omega}{d\phi} \right|
= P_\omega(R \sin\phi ) \cdot R \cos \phi.
\label{nphi_coh}
\qe

The rotors with $|\omega(\br)| > R$, on the other hand, 
form the incoherent group. The actual frequency of
a rotor belonging to this group is given by
\eq
\tomega_i = \f{2\pi }{\int_0^{2\pi} d\phi \f{dt}{d\phi}}
= \sqrt{\omega_i^2 - R^2}.
\qe
The phase distribution $n'(\phi) = n'(\phi; \omega_i)$
of an incoherent rotor is proportional to the frequency
it comes to $\phi$:
\eq
n'(\phi; \omega_i) = C |\dot{\phi}|^{-1} = C |\omega_i - R \sin \phi |^{-1}
\label{nphi_incoh}
\qe
with the normalization factor $C = \sqrt{\omega_i^2 - R^2}/2\pi$.

Now we replace the factor $e^{i\phi_j}$ in the RHS of eq. (\ref{ansatz})
by its ensemble average as
\eq
R = \sum_{\br'} G(\br-\br') \langle e^{i\phi} \rangle.
\qe
The average is the sum of the contributions from
the coherent and incoherent groups,
$
\langle e^{i\phi} \rangle_{coh} = \int_{-\pi/2}^{\pi/2} d\phi \, n(\phi) e^{i\phi}
$
and
$
\langle e^{i\phi} \rangle_{incoh} =
\int_{|\omega|>R} d\omega P_\omega(\omega) \int_{-\pi}^{\pi} d\phi n'(\phi; \omega)
e^{i\phi}.
$
The latter vanishes because of the symmetry of $P_{\omega}(\omega)$,
and the former with eq. (\ref{nphi_coh}) yields
\eq
R &=& R G_0 J(R),
\\
J(R)&=& \int_{-\pi/2}^{\pi/2} d\phi \, P_\omega(R \sin\phi) \cos^2 \phi,
\qe
with $G_0 = \sum_\br G(\br)$.
This is a self-consistent equation for the mean-field amplitude.
Expanding the integral as
$J(R)= (\pi/2) \left[P_\omega(0) + P''_\omega(0) R^2/8 + O(R^4)\right]$,
we obtain the critical coupling strength
\eq
G_{0c} = \f{2}{\pi P_\omega(0)}
\label{crit}
\qe
for the synchronization transition
(at which a non-vanishing solution $R$ appears).
For a square lattice,
we have $G_0 = 9.03 \cdot 3\gamma \omega_0/4\pi$.
Also we have $P_\omega(0) = 1/(\sqrt{2\pi} \omega_0 \delta_0)$.
Putting these together into eq.(\ref{crit}), we obtain
the critical angle
\eq
\delta_{0c} = 1.35 \gamma.
\qe
In the simulation we used $\gamma=0.1$,
which gives $\delta_{0c} = 0.135$ (rad) $= 7.73$ (deg).
This value is not very far from the numerically
obtained threshold angle $\delta_{th} = 6$ (deg).
Both the simulation and theory suggest that a quite 
narrow distribution of the mounting angle is required 
to achieve coordinated motion in a real bacterial carpet.
On the other hand, the mean-field theory predicts a sharp transition,
in contrast to the gradual crossover observed in the simulation.
Near $\delta_{0c}$, the orientational order parameter decays
as $S \propto \sqrt{\delta_{0c} - \delta_0}$, while the normalized
frequency deviation linearly approaches to the desynchronized limit
as $1- Q \propto \delta_{0c} - \delta_{0}$.
The distribution of the actual frequency $P_\Omega(\Omega)$
for $\delta<\delta_{0c}$ has three distinct peaks,
one at $\omega=0$ (the coherent group) and two symmetric peaks
for $\omega>0$ and $\omega<0$ (the incoherent group).
The central peak vanishes for $\delta>\delta_{0c}$.
These behaviors are qualitatively different from 
the numerical results shown in Figs. 2 and 3.
In order to explain the unconventional transition behavior, 
it is necessary to develop a theoretical framework 
that incorporates spatial fluctuations, which is 
an interesting problem for the future.

\paragraph{Conclusion. -}

The synchronization transition caused by long-range hydrodynamic
coupling is shown to be more gradual than for the global coupling.
The threshold angle for the crossover is estimated, and is not
far from the mean-field estimate for the transition point.
It suggests that a very narrow angle distribution is 
required to achieve correlated motion in a real bacterial carpet.
We believe that this work sheds some light on the delicate issues involved in
hydrodynamic synchronization and hope that it stimulates further theoretical
and experimental studies of the rich behavior of such systems.

\acknowledgments

NU thanks the hospitality at the University of Sheffield
where this work was initiated, and financial support
from Grant-in-Aid for Scientific Research from MEXT.
RG acknowledges financial support from the EPSRC.

\end{document}